\documentclass[preprint,nofootinbib,showpacs,preprintnumbers,pre,aps]{revtex4-1}
\usepackage{epsf, epsfig,graphicx}
\usepackage{subfigure}
\usepackage{bm}
\usepackage{amsmath}
\usepackage{amssymb}
\usepackage{color}

\begin{document}
\title{Dynamics of a membrane coupled to an active fluid}

\author{
  Chia-Chun Liang$^{1}$\footnote{Present address: Department of Earth System Science, University of California, Irvine, California 92697, USA}, Kento Yasuda $^{2}$, Shigeyuki Komura$^{2}$, Kuo-An Wu$^{1}$, and Hsuan-Yi Chen $^{3,4}\footnote{email address: hschen@phy.ncu.edu.tw}$
}
\affiliation{
  $^{1}$Department of Physics, National Tsing Hua University, Hsinchu 30013, Taiwan\\
  $^{2}$ Department of Chemistry,  Graduate School of Science, Tokyo Metropolitan University 192-0397 Tokyo, Japan \\
  $^{3}$Department of Physics, National Central University, Jhongli 32001, Taiwan\\
  $^{4}$Institute of Physics, Academia Sinica, Taipei, 11529, Taiwan
}	 	

\date{\today}

\begin{abstract}
	 
The dynamics of a membrane coupled to an active fluid on top of a substrate is considered theoretically.  
It is assumed that the director field of the active fluid has rotational symmetry in the membrane plane.  
This situation is likely to be relevant for \textit{in vitro} reconstructed actomyosin-membrane system.
 Different from a membrane coupled to a polar active fluid, this model predicts
that 
only when the viscosity of the fluid above the membrane is sufficiently large, a contractile active fluid is 
able to slow down the relaxation of the membrane for perturbations with wavelength comparable to the 
thickness of the active fluid.
Hence our model predicts a finite-wavelength instability in the limit of strong contractility, which is 
different from a membrane coupled to a polar active fluid.
On the other hand, a membrane coupled to an extensile active fluid is always unstable against long wavelength 
perturbations due to active extensile stress enhanced membrane undulation.
\end{abstract}	

\maketitle

\section{Introduction}

Investigating the dynamics of membranes is important both in the advance of cell 
biophysics~\cite{ref:Sackmann_book, ref:Lodish_book} and soft condensed matter 
physics~\cite{ref:Seifert_1997}.
An interesting example that bridges the equilibrium soft matter, nonequilibrium active matter, and biophysics, is the flickering phenomena 
of red blood cells. 
It was found that the fluctuation power spectrum of the membrane depends on the  viscosity of the solvent~\cite{ref:Tuvia_97}, 
and the fluctuation-dissipation relation is violated due to active processes in the red blood cells~\cite{ref:Betz_2016,ref:Komura_2016}.  
Membranes that contain active inclusions are also nonequilibrium soft matter. 
A membrane containing active pumps have fluctuation power spectrum that depends on the active force dipoles exerted by 
the pumps~\cite{ref:Prost_2001}, and instabilities can occur when the membrane curvature enhances the activities of the 
pumps~\cite{ref:Prost_2000}.  
Furthermore, it has been predicted that finite size domain can be induced due to the active conformational transitions of the
 inclusions~\cite{ref:Chen_2004, ref:Chen_2010}.

In red blood cells, cell membranes are coupled to a relatively regular spectrin network~\cite{ref:Sackmann_book}.  
In most biological cells, on the other hand, the cell membranes are coupled to the actomyosin cytoskeleton~\cite{ref:Lodish_book}, 
in which actin filaments are densely cross-linked and myosin motors exert contractile stress~\cite{ref:Prost_2007}.  
The spatial organization of actin filaments and the way that contractile stress is coupled to the membranes are different 
from red blood cells and membranes containing active inclusions.  
To study how membrane dynamics are influenced by the out-of-equilibrium cytoskeleton, we propose in this article  a minimal 
model in which the membrane is in contact with a passive simple fluid on one side, whereas the other side is filled with 
an active fluid on a solid substrate.
The direction of contractile stress is dictated by the orientation of the actin filaments which we assume to be distributed isotropically 
in the plane that is parallel to the membrane.
Thus our model describes a fluid membrane on top of an active fluid which could be reconstructed \textit{in vitro}~\cite{Tsai}, or 
seen \textit{in vivo} but away from the relatively highly polarized lamellipodium.  
The generality of this model also allows it to describe the surface dynamics of a fluid containing either contractile or extensile 
non-swimming rod-shaped active particles on a solid substrate~\cite{ref:Paxton_2004}.
However, it is important to notice that the system under consideration in our model is slightly different from previous studies of 
the dynamics for active fluid thin film~\cite{Ramaswamy, Basu}.
In~\cite{Ramaswamy,Basu}, active particles with polar order were considered and the focus was the coupling effect between membrane dynamics 
to the director fluctuations in the membrane plane.

The main result of our calculation shows that, for contractile active fluids,  active stress generally helps to stabilize 
a perturbed membrane.  
However, when the passive fluid is sufficiently viscous, the contractility in the active fluid can slow down perturbations 
with wavelengths comparable to the thickness of the active fluid.  
This means that at sufficiently strong contractility, the membrane can undergo a finite-wavelength instability. 
Similar to many active polar or nematic systems~\cite{Ramaswamy2002}, the origin of this instability is the 
coupling between the splay deformation of the director field and the active stress generated flow.  
However, the fact that the above instability occurs at wavelengths close to the thickness of the active fluid marks 
a significant difference from the long-wavelength instability predicted for a membrane on top of 
a thin active polar fluid film~\cite{Ramaswamy,Basu}, and the active Fr\'{e}edericksz transition~\cite{Freedericksz}.  
For extensile active fluids, we further find that the system is always unstable in the long wavelength limit due to the 
same coupling between director and flow field.  
The membrane dynamics presented in our model is valid from thin film limit, where lubrication approximation holds, to 
bulk limit where wavelength is small compared to the thickness of active fluid.  
Extension of our model to polar active fluid and systems with polymerization/depolymerization dynamics will be interesting.

This article is organized as following.  
In Sec.~II we present our model from the geometry of the system, the elasticity of the director field and membrane, 
to the equations of motion and boundary conditions.  
Sec.~III presents the membrane relaxation rates calculated from our model.  
Also presented and discussed are phase diagrams showing the parameter range where the system becomes unstable.  
In Sec.~IV we compare our model with previous studies on similar systems~\cite{Ramaswamy,Basu,Freedericksz}, 
and discuss possible experimental realizations.
Some details on the calculation in our model and the solution in the thin film limit are reported in the Appendices.

\section{Membrane on top of an active viscous fluid}

\begin{figure}[tbh]
\begin{center}
\includegraphics[scale=0.3]{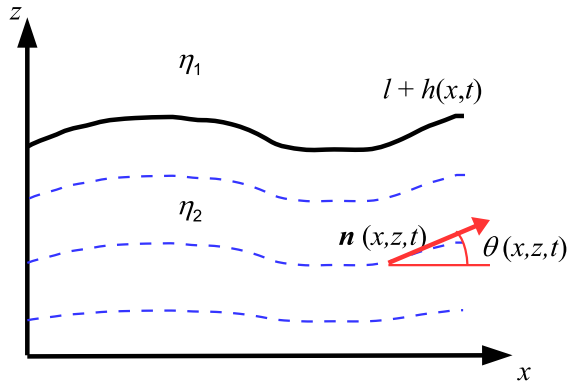}
\end{center}
\caption{Schematics of the system.  A simple fluid of viscosity $\eta _1$ is above the membrane, below the 
membrane is an active viscous fluid with viscosity $\eta _2$ and director field $\mathbf{n}$.  The height of the membrane 
is $l$.  
The dashed lines represent the director field of the active fluid. The angle between local director field and the $xy$ plane is denoted as 
$\theta$.   
A solid surface is located at the $xy$ plane.  For simplicity we consider perturbation from flat membrane state with a 
single wave vector $q \hat{\bf x}$, and the system is translationally invariant in the $y$-direction.
}
\label{f:schematics}
\end{figure}

Consider an active fluid on a solid substrate, 
the upper surface of the active fluid is a membrane whereas the upper side of the membrane is filled with a simple fluid, 
as represented in Fig.~\ref{f:schematics}. 
In the steady state, the membrane is flat at $z = l$, in the presence of fluctuations the membrane position becomes
$z(x,y,t)=l + h(x,y,t)$.  

We assume that the director field of the active fluid is parallel to the membrane due to surface anchoring effect but otherwise isotropic. 
This means that the director field is in the isotropic phase and thickness of the active fluid is not very large.   
When this system is in the steady state, the tensor $\langle n_i n_j \rangle$ vanishes for all $i$ and $j$  
except $\langle n_x n_x \rangle=\langle n_y n_y \rangle=1/2$. 
Here $n_i$ is the $i$-th component of the unit vector $\mathbf{n}$ indicating the local director direction, 
and $\langle n_i n_j \rangle $ is the average of $n_i n_j$ over a small region large compared to molecular size.   
This geometry mimics the situation for an artificial membrane on top of a reconstructed actin-myosin network above 
a solid surface. 
However, it is different from the lamellipodium of a cell, in which the organization of actin filaments is 
polarized~\cite{Prost_2015}.  
If we replace the membrane by an interface between the active and the passive fluids, this model can describe 
a fluid containing non-swimming rod-shaped active particles that follow the plane of the interface but 
without in-plane nematic order.  

When the membrane is deformed, there is an elastic energy associated with the deformation. 
The elastic free energy of the membrane can be expressed as 
\begin{eqnarray}
F =  \frac{1}{2}\int dxdy  \, \left[ \gamma(\nabla h)^2 + \kappa(\nabla^2 h)^2 \right], 
\end{eqnarray}
where  $\gamma$ is the membrane tension and $\kappa$ is the bending rigidity.

Without loss of generality, we consider a perturbation on the membrane with wave vector $q \hat{\mathbf{x}}$, and assume 
translational invariance in the $y$-direction.
Then the membrane deformation $h(x,y,t)$ can be written as  
\begin{eqnarray}
h(x,y,t) = h(q,t) e^{i q x} + c.c.,
\end{eqnarray}
where $c.c.$ is the complex conjugate of $h(q,t) e^{i q x}$. 
Let us denote the angle between the local director field of the active fluid and the $xy$-plane 
by $\theta$.  
The symmetry of the system indicates that the elastic free energy of the director field is
\begin{eqnarray}
F_d = \int dxdydz \ \left\{ 
     \frac{k_{\perp}}{2}\left[     \left(\frac{\partial \theta}{\partial x}\right)^2
                                               + \left(\frac{\partial \theta}{\partial y}\right)^2 \right]
                                      + \frac{k_z}{2} \left(\frac{\partial \theta}{\partial z}\right)^2 \right\},
\label{eq:Fd}
\end{eqnarray}
where $k_{\perp}$ and $k_z$ are elastic constants analogous to the the splay and bending constants of nematic liquid crystal~\cite{ref:LC_book}.  
Notice that the symmetry of our system does not allow the system to acquire elastic energy against twist deformation.  For simplicity we assume that the director field relaxes much faster than other slow modes.
Use $\theta \sim e^{iqx}$ and $\partial \theta /\partial y=0$, minimize $F_d$ with respect to $\theta$, 
we obtain the equation for the angle $\theta$ in the active fluid in the limit of small $\theta$ as 
\begin{eqnarray}
k_{z} \frac{d^2 \theta}{dz^2} - k_{\perp} q^2 \ \theta =0.
\end{eqnarray}

Let the director field on the boundaries be parallel to the membrane and the solid surface, in the limit of small membrane deformation, 
$\theta (z=l,t) = i q h(q,t)$ 
and $\theta (z=0,t) = 0$.
Using one-constant approximation $k_{\perp} = k_z $, we obtain the following director field in the active fluid
\begin{eqnarray}
\theta (z,t) = i q h(q,t) \frac{\sinh(q z)}{\sinh(q l)}.
\end{eqnarray}
 
The actomyosin network is driven away from equilibrium by several active processes.  
First, actin filaments grow by recruiting free actin monomers to the plus ends of the filaments, and this is balanced by 
actin depolymerization in the minus ends of the filaments.   
Second, myosin motors utilize energy of ATP hydrolysis to exert forces to the actin filaments and surrounding solvent.
They provide the cytoskeleton a contractile stress that depends on local orientation of actin filaments and myosin density.  
In this article, we focus on the effect of active contractility, while actin polymerization/depolymerization is neglected.
Moreover, the density of actin network in the system is treated as a constant for simplicity.

The local active stress in the active fluid is given by~\cite{ref:Prost_2007}
\begin{eqnarray}
\sigma^{\rm a}_{ij} = \chi \tilde{Q}_{ij},
\end{eqnarray}
where $\chi >0$ ($\chi <0$) for a contractile~\cite{ref:Bray_book} (extensile~\cite{ref:Wu_PNAS}) active fluid, and $\tilde{Q}_{ij}$ is the nematic order parameter, i.e., the 
traceless part of $\langle n_i n_j \rangle $. 
Although there is also an isotropic active stress, it does not affect the dynamics of the system because of the incompressibility 
condition imposed on the active fluid.  
In the limit of small deformation, the relevant components of $\tilde{Q}_{ij}$ are
 \begin{eqnarray}
\tilde{Q}_{xz}=\tilde{Q}_{zx}=\theta /2.
\end{eqnarray}

The dynamics of the membrane is coupled to the surrounding fluids.  
For $z > l$, there exists a passive fluid with viscosity $\eta_1$.   
The dynamics of this passive fluid is described by the Stokes equation with incompressibility condition
\begin{eqnarray}
\label{eq:region_1}
 \eta _1 \nabla^2 \mathbf{v}-\nabla p  + \mathbf{f}_1  \delta (z-l)=0, 
 \label{eq:momentum1}
 \end{eqnarray}
and
\begin{equation} 
\nabla \cdot \mathbf{v}=0,
\end{equation}
where $\mathbf{v}$ is the flow field, $p$ is the pressure, and $\mathbf{f}_1$ is the force exerted on the fluid by the membrane~\cite{ref:Seifert_1997}.

Between the membrane and the rigid surface, there exists an active fluid with viscosity $\eta_2$.
In this active fluid, actin filaments are crosslinked thanks to the reversible bonds due to various linkers.  
This gives the network a mechanical response that is elastic on short time scale, but viscous on long time scale.  
We are interested in the dynamics on lengths greater than the typical molecular size in the active fluid ($\sim 10 \ {\rm nm}$),  
and long time limit in which the cytoskeleton between the membrane and the rigid substrate can 
be modeled as a viscous fluid under the influence of active forces.
 In this limit the active fluid is described by the following equation
\begin{eqnarray} 
\label{eq:region_2}
\eta_2\nabla^2 \mathbf{v} -\nabla p   + \mathbf{f}^{\rm a} +\mathbf{f}_{2}^{\rm u} \delta (z-l) 
+ \mathbf{f}_{2}^{\rm d} \delta (z)=0, 
\label{eq:momentum2}
\end{eqnarray}
where
\begin{eqnarray}
f^{\rm a}_i = \chi \partial _j \tilde{Q}_{ij}
\end{eqnarray}
comes from the active stress, $\mathbf{f}_2^{\rm u}$ and $\mathbf{f}_2^{\rm d}$ are the forces from the membrane and the 
supporting substrate, respectively.  
Similar to the passive fluid above the membrane, we assume that the active fluid is incompressible.

In our model, the membrane is treated as a two-dimensional surface with zero thickness and 
the following conditions need to be satisfied at this surface.  
First, the flow field needs to be continuous, 
\begin{eqnarray}
\mathbf{v}(z = l^+) = \mathbf{v}(z = l^-).
\label{eq:bc12}
\end{eqnarray}
Second, the shear stress is continuous across the membrane because the membrane is a two-dimensional fluid,
\begin{eqnarray}
\sigma _{xz}(z= l^+) = \sigma_{xz}(z= l^-).
\label{eq:bc03}
\end{eqnarray}
Finally, the discontinuity of normal stress across membrane balances with the elastic force of the membrane,
\begin{eqnarray}-\sigma_{zz}(z= l^+) \ + \sigma_{zz}(z= l^-)=-\frac{\delta F}{\delta h}.
\label{eq:bc04}
\end{eqnarray}
At the rigid surface $z=0$, on the other hand, we assume the no-slip condition for the flow field,
\begin{eqnarray}
\mathbf{v}(z=0)=0.
\label{eq:bc56}
\end{eqnarray}

\section{Linearized hydrodynamics close to the steady state}

With incompressibility condition, the velocity and pressure in the system can be expressed in terms of the membrane conformation 
and the boundary forces $\mathbf{f}_1$, $\mathbf{f}_2^{\rm u}$ and $\mathbf{f}_2^{\rm d}$.   
These forces are further solved by imposing the boundary conditions Eqs.~(\ref{eq:bc12})--(\ref{eq:bc56}).  
Time evolution of the membrane height is obtained from the kinematic boundary condition $v_{z}|_{z = l} = \partial h/\partial t$.  
Detailed intermediate steps of this calculation are presented in Appendix~A.

The resulting equation for membrane dynamics is
\begin{eqnarray}
\frac{\partial h(q,t)}{\partial t}
&= - \left[\lambda _{\rm p}(q)+\lambda_{\rm a}(q) \right]  h(q, t),
\label{eq:dhdt}
\end{eqnarray}
where
\begin{eqnarray} 
\lambda_{\rm p}(q)  = 
\frac{(\gamma q + \kappa q ^3 )[-1+E + e^{4 q l}(1+E) - 2 e^{2 q l} (E+ 2 q l+2 E q^2 l^2)]}
{2 \eta _2   \left[ (-1+E)^2 + e^{4 q l}(1+E)^2 - 2 e^{2 q l} (-1+E)(1+2 q^2 l^2) \right]}
\label{eq:lambdap}
\end{eqnarray}
is the contribution from the restoring force provided by membrane elasticity.
In the above, $E  = \eta _1 /\eta_2$ specifies the viscosity ratio of the passive fluid at $z>l$ and the active fluid at $z<l$.  On the other hand, 
\begin{align}
\lambda_{\rm a}(q)& =
\frac{\chi}{32 \eta _2 \sinh (q l) \left[-q^2 l^2 (-1+E ^2)+ \{\cosh(q l)+E \sinh(q l)\}^2 \right]}
\nonumber \\
\times & \left[ (1+ 4 q^2 l^2 - 4 E q ^3 l^3)\sinh(q l)-(E + 4 q l +4 E q^2 l^2 + q ^3 l^3)\cosh(q l) 
\right.
\nonumber \\
& \left. +\sinh (3 q l)+E \cosh (3 q l ) \right]
\label{eq:lambdaa}
\end{align}
comes from the active stress associated with spatially nonuniform director field.

In the limit $q l  \gg1$ the relaxation rate approaches
\begin{align}
\lambda_{\rm p} &=\frac{\gamma q +\kappa q^3}{2(\eta_1 + \eta _2)},
\\
\lambda_{\rm a} &= \frac{\chi}{8 (\eta _1 + \eta _2)}. 
\label{eq:largeq}
\end{align}
The passive contribution reduces to that for a membrane separating two bulk fluids. 
The active contribution is independent of $q$, and it 
indicats that when $ql \gg 1$ a contractility assisted membrane relaxation is faster 
than a passive system.  On the other hand, for an extensile active fluid, 
active stress stretches a curved membrane and prefers a curved membrane configuration, 
and hance the membrane relaxation is slower than a passive system~\cite{footnote}. 
Note that at sufficiently large $q$ (for a tensionless membrane this happens when $q \gg (\chi /\kappa)^{1/3}$), 
the passive contribution dominates the membrane decay rate and the membrane is basically passive.

In the limit $q l \ll 1$, on the other hand, the relaxation rate is 
\begin{align}
\lambda _{\rm p} &\approx \frac{\gamma q + \kappa q ^3}{\eta _2} \left( \frac{q^3 l^3}{3}- \frac{\eta_1}{2 \eta_2}q^4 l^4
+ \cdots \right),
\\
\lambda_{\rm a} &\approx \frac{\chi q^2 l^2}{12 \eta _2}\left( 1-3\frac{\eta_1}{\eta_2} q l + \cdots \right). 
\label{eq:smallq}
\end{align}
Here the leading contributions are independent of the viscosity of the passive fluid as expected from the lubrication
approximation~\cite{ref:Bachelor}, whereas the contributions from the passive fluid ($\eta_1$) appear only in the next 
order in $q l$.   
It is worth noticing that when $q l$ is small, active stress dominates the membrane relaxation, 
$\lambda_{\rm p} < \lambda_{\rm a}$, and the system is stable (unstable) when the active fluid is contractile, 
$\chi>0$ (extensile, $\chi<0$).  
The mechanism for this long-wavelength instability in extensile system is again due to active stretching induced  
membrane undulation~\cite{ref:Maitra_2014}, which is also responsible for the sign of $\lambda _{\rm a}$ in Eq.~(\ref{eq:largeq}).

\begin{figure}[tbh]
	\mbox{
	  \subfigure[]{\epsfig{figure=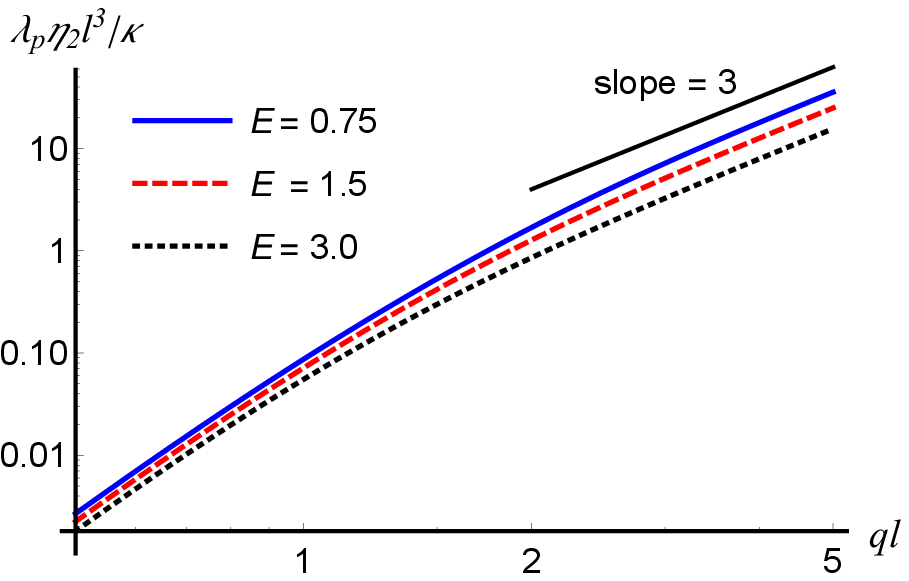, width=3 in}}	
	}
	\mbox{
	  \subfigure[]{\epsfig{figure=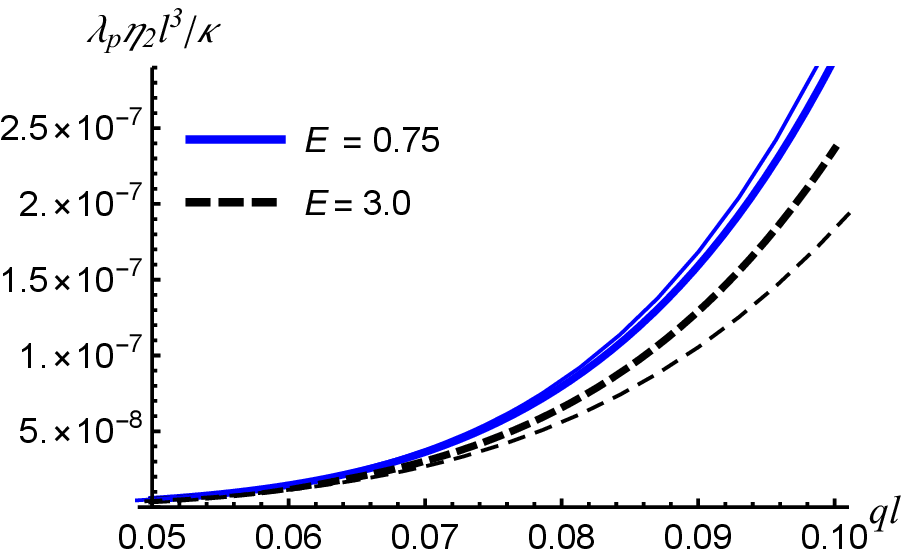, width=3 in}}	
	}
\caption{(a) The contribution of membrane elasticity to membrane relaxation rate for a tensionless membrane.  In the limit of large $ql$ it scales as $q^3l^3$. Curves for $\eta_1/\eta_2= E= 0.75$ (solid), $1.5$ (long dashed), and $3.0$ (short dashed) are shown.  
(b) Contribution of membrane elasticity to membrane relaxation rate for a tensionless membrane in the long wavelength limit.  Solid curves: $E = 0.75$, dashed curves: $E = 3.0$.  Thin curves represent the long wavelength expansion presented in Eq.~(\ref{eq:smallq}), thick curves represent the full solution presented in Eq.~(\ref{eq:lambdap}).
}
\label{f:prelaxation_rate}
\end{figure}

\begin{figure}[tbh]
\mbox{
	  \subfigure[]{\epsfig{figure=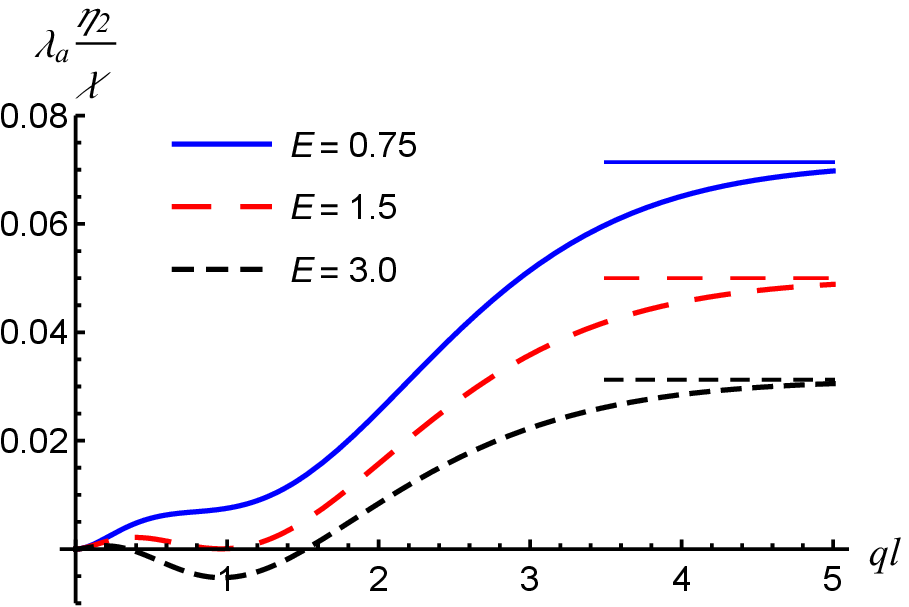, width=3 in}}	
	}
	\mbox{
	  \subfigure[]{\epsfig{figure=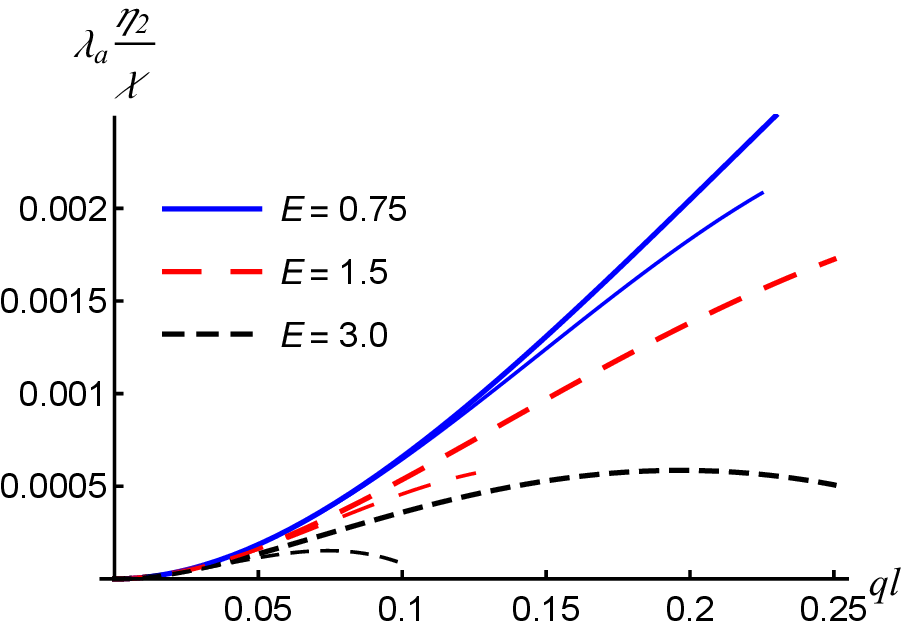, width=3 in}}	
	}
\caption{(a)  Contribution of contractility to membrane relaxation rate.
In the limit of large $q l$ the relaxation rate approaches that provided by the short-wavelenth limit solution in Eq.~(\ref{eq:largeq}).   Solid curves: $E = 0.75$, dashed curves: $E = 1.5$, dotted curves: $E = 3.0$.  Thick curves represent the full solution presented in Eq.~(\ref{eq:lambdaa}), thin lines represent the large $q l$ limit presented in Eq.~(\ref{eq:largeq}).  Notice that when $E>E_{\rm c} \approx 1.5$ there is a region where $\lambda _{\rm a}$ becomes negative.  
(b) Contribution of contractility to membrane relaxation rate in the long wavelength limit.  Thick curves represent the full solution presented in Eq.~(\ref{eq:lambdaa}), thin curves correspond to the results from the long wavelength limit expansion Eq.~(\ref{eq:smallq}).
}
\label{f:arelaxation_rate}
\end{figure}

The stability of the membrane in the long wavelength limit predicted by our model is different from a membrane on top 
of a thin polar active fluid film.  
In the latter case, a long-wavelength instability due to the coupling between the fluctuations of the director field in the 
membrane plane (the $xy$ plane in our geometry) and membrane deformation was predicted~\cite{Ramaswamy, Basu}. 
This instability can be stabilized by the active force quadrupoles of the active fluid in the thin film limit~\cite{ref:Lenz_PNAS_2018}.  
In our model, such a coupling does not exist because the system does not have any in-plane polar order.
In Refs.~\cite{Ramaswamy,Basu}, a long wavelength instability associated with splay of the director field 
in the $xz$ plane for a contractile active fluid is also discussed, which is again different from our prediction.  
To discuss this difference in details, we present the lubrication approximation of our model in Appendix~B by keeping track of 
the order of magnitude of the sub-leading terms.  
There we show that a consistent expansion at $q l \ll 1$ term by term  leads to a stable (unstable) membrane in the long 
wavelength limit when the membrane is coupled to a contractile (extensile) active fluid.

To see the crossover from short wavelength to long wavelength behavior, we plot in Figs.~\ref{f:prelaxation_rate} and 
\ref{f:arelaxation_rate} the passive and active contribution to the membrane relaxation rate as functions of $q l$ for 
a tensionless membrane ($\gamma=0$) with different $E$.  
Figure~\ref{f:prelaxation_rate} shows that the passive contribution increases monotonically with $q$ due to the 
bending free energy density $\kappa q^4/2$.  
In the limit of large $q l$, this part scales as $q^3 l^3$ as expected.  
In the limit of small $q l$, the long wavelength expansion Eq.~(\ref{eq:smallq}) holds reasonably well as $q l$ becomes 
sufficiently small.
Figure~\ref{f:arelaxation_rate} shows that the active contribution in the large $q l$ limit also approaches the prediction 
provided by Eq.~(\ref{eq:largeq}).  
In the long wavelength limit, the expansion presented in Eq.~(\ref{eq:smallq}) again holds better as $q l$ becomes sufficiently 
small.  
For both passive and active contributions, the long wavelength expansion agrees better with the full solution when 
$E = \eta _1 / \eta _2$ is smaller.

\begin{figure}[tbh]
\mbox{
	  \subfigure[]{\epsfig{figure=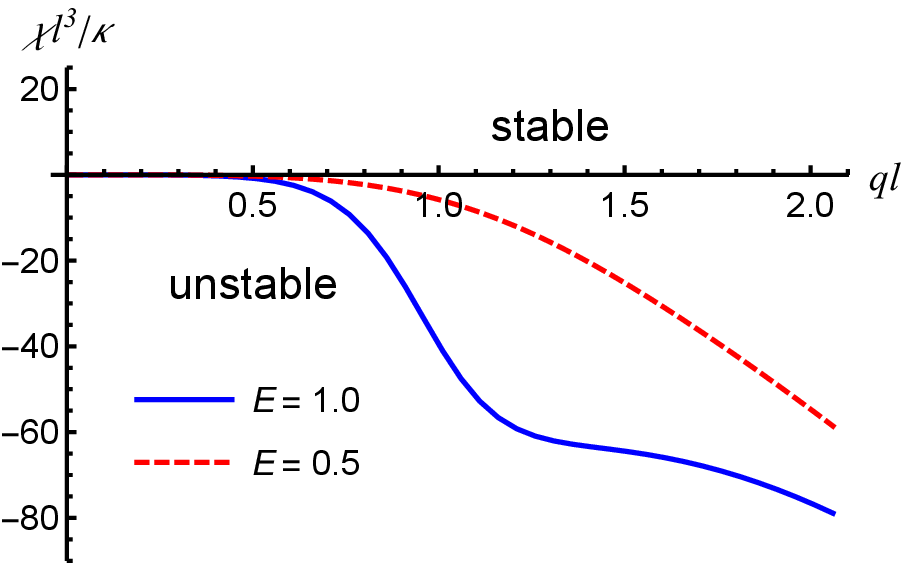, width=3 in}}	
	}
	\mbox{
	  \subfigure[]{\epsfig{figure=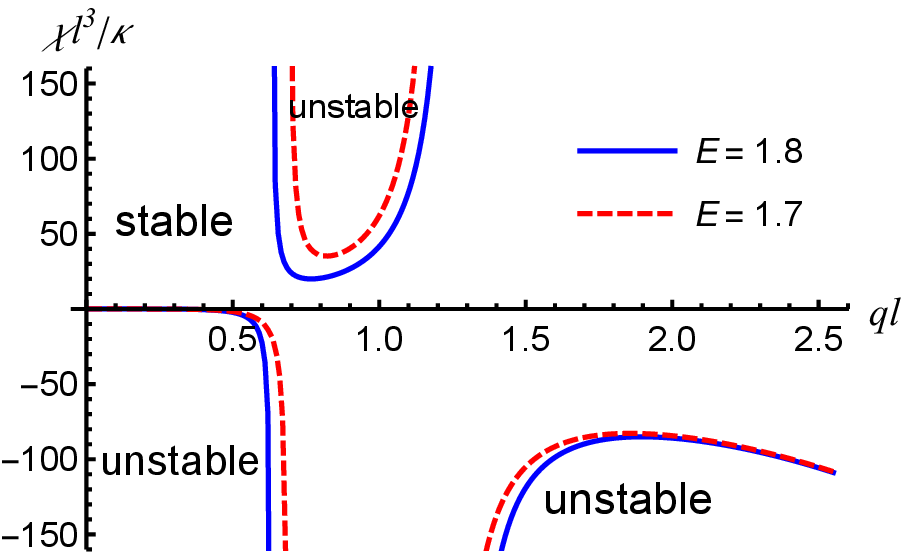, width=3 in}}	
	}
\caption{(a)  For $E < E_{\rm c}$  (blue solid curve: $E = 1.0$, red dashed curve: $E = 0.5$, $E_{\rm c} \approx 1.5$), the system becomes unstable when the active fluid is extensile and activity is sufficiently strong.  Region below the curves is where the system becomes unstable.  In the limit of small $ql$ the system is always unstable if the active fluid is extensile.  
(b) For $E>E_{\rm c}$, (blue solid curve: $E = 1.8$, red dashed curve: $E = 1.7$), there is a finite wavelength instability if the active fluid is contractile.  If the active fluid is extensile, in the small $ql$ limit the system is always unstable;  $ql > 1$ there is also an finite-wavelength instability.   
}
\label{f:phase_diagram}	
\end{figure}

\begin{figure}[tbh]
\begin{center}
\includegraphics[scale=0.6]{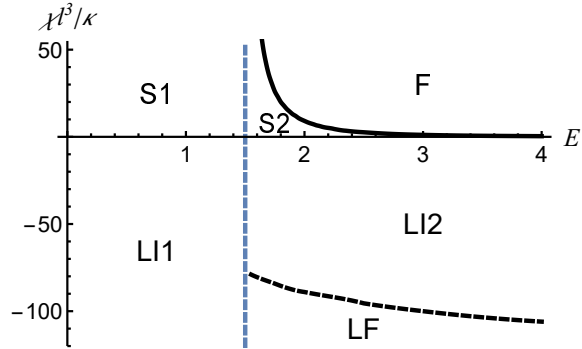}
\end{center}
\caption{Complete phase diagram for the system. The vertical dashed line indicates $E=E_{\rm c} \approx 1.5$.  
In regions S1 and S2, the system is stable.
In region LI1 and LI2, the system has a long-wavelength instability.
In region F, there is a finite wavelength instability.  
In region LF, there is a long-wavelength instability and a finite-wavelength instability. 
}
\label{f:phase_diagram_ext}	
\end{figure}

The active contribution also show interesting features that depends on the value of $E$, which 
can be clearly seen from Fig~\ref{f:arelaxation_rate}(a).  
For small $E$, the active contribution also increases monotonically with $q$ until it reaches the limiting 
magnitude provided by Eq.~(\ref{eq:largeq}), indicating contractility-assisted membrane relaxation at all wavelengths.  
However, as $E$ becomes larger than $E_{\rm c} \approx 1.5$, contractility slows down relaxations for perturbations 
with $q l \sim 1$.  
This indicates that even though contractility assists membrane relaxation in the short- and long-wavelength limits, 
there is a possible contractility-induced membrane instability for an active fluid with sufficiently strong active stress
when $E > E_{\rm c}$.  
In contrast, for a membrane on top of an extensile active fluid, $\chi < 0$, the system is always unstable 
in the long wavelength limit regardless the magnitude of membrane tension, as can be seen from Eq.~(\ref{eq:smallq}).

The stability diagram for a membrane on top of an active fluid can be plotted by introducing a dimensionless parameter 
$\chi l^3 /\kappa$ which characterizes the relative strength of the active stress with respect to membrane elasticity.   
It can be seen from Fig.~\ref{f:phase_diagram}(a) that when $E<E_{\rm c}$, a system with a contractile active fluid ($\chi >0$) is always stable.  On the other hand, when the active fluid is extensile ($\chi <0$), the membrane is always unstable in the long wavelength limit.     
Figure~\ref{f:phase_diagram}(b) shows that when $E>E_{\rm c}$, there is a finite-wavelength instability when the active fluid is contractile ($\chi >0$), and the first unstable mode has $ql$ of order unity.  When the active fluid is extensile ($\chi <0$), the system is always unstable in the short wavelength limit, and there is another finite-wavelength instability in the negative $\chi$ region when the activity is sufficiently strong.  Two important features in $E>E_{\rm c}$ regime should be noticed.  First, due to limited range of $ql$ that can be shown in the figure, it is not easy to see that in the short wavelength limit the system is always stable because membrane elasticity is the dominating drive of the relaxation dynamics of the system.  Second, an extensile system is stable in the wavelength regime where a contractile system has a finite-wavelength instability.  This is simply because $\lambda$ changes sign when $\chi$ changes sign.     
The complete phase diagram which summarizes the above discussion of the stability of the system is shown in Fig.~\ref{f:phase_diagram_ext}.  

\section{Discussion}

The instability for a membrane on top of a contractile active fluid is similar to the active Fr\'{e}edericksz transition for 
active nematics~\cite{Freedericksz} because in both cases the instability is a result of the coupling between splay 
deformation and flow field.  
However, in active Fr\'{e}edericksz transition, the upper boundary does not deform and the maximum deformation 
of the director field happens not at the upper boundary of the active fluid but in the middle.  
When the transition occurs, the system develops a spontaneous flow that is independent of $x$.
Furthermore, for this transition to occur, the director field needs to have a finite relaxation rate.

On the other hand, our model focuses on the limit of fast director relaxation.
With a membrane that can deform on the top, in our model the maximum deformation of the director field occurs on the upper boundary of the active fluid.  
Thus the instability in our model manifests itself as a membrane undulation instability, and it strongly depends on 
the viscosity of the passive fluid. 
By comparing Ref.~\cite{Freedericksz} with our model, we expect that for a system with a sufficiently viscous passive fluid ($E > E_{\rm c}$) 
and relatively slow director dynamics, active Fr\'{e}edericksz transition should happen first; but for a system with 
$E > E_{\rm c}$ and fast director relaxation,  
one should first observe membrane undulation instability. 
For a system with a passive fluid that is less viscous ($E < E_{\rm c}$), 
active Fr\'{e}edericksz transition may be observed, but not finite-wavelength membrane undulation instability.

Previous studies of thin active fluid film applied lubrication approximation to obtain the coupled dynamics of 
membrane fluctuations and  the deformation of polar order in the membrane plane~\cite{Ramaswamy, Basu}. 
Recently, it has been pointed out in Ref.~\cite{ref:Lenz_PNAS_2018} that the next order active force multipole 
can restore the stability of the system considered in Refs.~\cite{Ramaswamy, Basu}.   
Although these results are important progress in two-dimensional active nematics, their main focus was the 
dynamics of the director in the plane parallel to the membrane.  
In this aspect, the focus of these studies is not related to our model.

It is interesting to point out that in both Refs.~\cite{Ramaswamy,Basu}, the 
possible long wavelength instability for director splay in the $xz$ plane are also discussed.  
This is closely related to our model, but their prediction is different from our result for $q l \ll 1$.
To find out the origin of this difference, we present in Appendix~B a small-$q l$ expansion and show that, 
for a membrane on top of a contractile active fluid, the system is stable in the long wavelength limit if one 
consistently takes all terms of lowest order in $q l$ into account.  
The finite wavelength instability predicted by our model appears only when higher order terms in small $q l$ 
expansion are included.
These terms depend on the viscosity of the passive fluid and are clearly beyond the regime where lubrication 
approximation holds.  
This further shows that our general theory is a good approach to study the membrane dynamics coupled to 
an active fluid, ranging from $q l \ll1 $ to $q l \gg 1$ limits.

In typical {\it in vitro} experiments, the viscosity of the passive fluid is at best as large as the viscosity of the 
active counterpart.
Our model predicts that in these systems the membrane relaxation is faster 
(slower) than a passive system if the active fluid is contractile (extensile).  
To observe a contractility-induced finite-wavelength instability, one may add passive polymers to the passive 
fluid to increase $E$.  
For $E  \approx 1.7$ -- $2.0$, finite wavelength instability can be observed for $\chi l^3 /\kappa \ge 10$ -- $30$.

The parameter $\chi$ has dimension of energy per unit volume, since the contractility comes from the force 
dipoles of the molecular motors in the active fluid, the magnitude of $\chi$ is of the order of the typical magnitude 
of single force dipole times the density of the motors.  
For a typical biomembrane $\kappa \approx 20$ -- $40$~$k_{\rm B} T$~\cite{ref:Sackmann_book}, and the 
magnitude of a typical force dipole generated by a molecular motor is around $10$~$k_{\rm B} T$,
the dimensionless parameter $\chi l^3/\kappa$ has magnitude around $c_0 l^3$ where $c_0$ is the average 
number density of  molecular motors.  
Since we typically have $c_0^{-1/3} \ll l$, it is easy for the instability to be observed.

We assume the director field to be isotropic in the membrane and substrate planes. Therefore the
active fluid is in an isotropic phase in the bulk, and the orientational order is due to surface anchoring.
In real experiments, such spatial director distribution can be achieved when the thickness of
active fluid is not very large compared to the nematic correlation length (roughly, a few times the ``mesh size'' of the network formed by the active particles).  
For example, in vitro actyomyosin systems have mesh size at least on the order of $\sim 50 \ {\rm nm}$, and the director field can be prepared to be parallel to the substrate for active fluid with thickness up to $\sim 1 \ \mu$m.  
For non-motile rod-shaped bacteria, the thickness of the active fluid can be as large as $\sim 10 \ \mu {\rm m}$.  
This means that, first, it is appropriate to apply hydrodynamic theory to describe our model systems.  
Furthermore, Fig 2a and Fig 3a show that, even when $ql \sim 5$, the large $ql$ expressions Eqs.~(19) and (20) already describe the membrane dynamics well.  
Hence our analysis, even the large $ql$ limit, is experimentally accessible.

In summary, we have studied the dynamics of a membrane above a finite-thickness active fluid beyond lubrication 
approximation.  
Our model predicts that, in the short wavelength limit ($q l \gg 1$, $q \gg (\chi /\kappa)^{1/3}$), the 
membrane relaxation is dominated by the passive driving of membrane elasticity.  
For longer wavelengths, the system is affected by active stress.  
For extensile active fluid, the membrane is always unstable in the long wavelength limit.
For contractile active fluid, the membrane can have a finite wavelength instability when the passive fluid above the 
membrane is sufficiently viscous and the strength of active stress exceeds a threshold that depends on $E$.  
The long wavelength instability predicted for extensile active fluid and finite wavelength instability predicted for the 
contractile active fluid are both within typical experimental parameter range.

In this study  we considered a relatively simple active fluid in which the density of active particles is uniform and the
director field is isotropic in the membrane plane.  
It will be very interesting to extend current model,  for example, by including the dynamics of director field, one can 
obtain a phase diagram which includes the membrane undulation instabilities and active Fr\'{e}edericksz transition.  
By considering situations where concentration variations in the system becomes important, especially when 
polymerization/depolymerization~\cite{Prost_2015} (or birth/death) of the active particles contribute significantly 
to the dynamics of the system, one can explore the physics of cytoskeleton (or biofilm). 
Finally, systems with different membrane-director coupling, 
for example, the situation when the director field is perpendicular to the membrane~\cite{ref:Basu_2}, 
and the active fluid has finite thickness, is also interesting and worth studying in the future.

\section*{Acknowledgments}

H.-Y. C. thanks J.-F. Joanny for interesting discussions.   
C.-C. L. thanks the hospitality and the financial support of TMU from NTHU-TMU cotutorial program.  K.Y.\ acknowledges support by a Grant-in-Aid for JSPS Fellows (Grant No.\ 18J21231) from the Japan Society 
for the Promotion of Science (JSPS). 
S.K.\ acknowledges support by a Grant-in-Aid for Scientific Research (C) (Grant No.\ 18K03567 and
Grant No.\ 19K03765) from the JSPS.
H.-Y. C. is supported by the Ministry of Science and Technology, Taiwan (Grant No. MOST 107-2112-M-008-024- and MOST 108-2112-M-008-016 ) and K.-A.W. acknowledge the support of Ministry of Science and Technology, Taiwan (Grant No. MOST 105-2112-M-007-031-MY3 and 108-2112-M-007-024-). H.-Y.C. and K.-A.W. acknowledges the support from National Center for Theoretical Sciences, Taiwan.   
The authors also thank an anonymous referee for the helpful suggestions on Eq.~(3).

\appendix
\section{Derivation of the membrane evolution equation}

To obtain the membrane evolution equations~(\ref{eq:dhdt}), (\ref{eq:lambdap}) and (\ref{eq:lambdaa}), we first use 
the incompressibility condition to obtain the pressure.  
This is done by taking the Fourier transform of the divergence of the momentum equations~(\ref{eq:momentum1}) and 
(\ref{eq:momentum2}).
The Fourier transform of the velocity field can be solved by substituting the pressure back to the Fourier transform 
of the momentum equations.

Next we take the inverse Fourier transform in the $z$-direction to obtain the pressure and flow field in terms of the boundary 
forces and active force.
The resulting expression for the pressure in the region $z>l$ is
\begin{eqnarray}
p(q,z,t) = \frac{e^{-q(z-l)}}{2}\left[ -i f_{1x}(q,t)+f_{1z}(q,t)\right],
\end{eqnarray}
and that in the region $z<l$ is
\begin{align}
p(q,z,t)
&=- \frac{\chi q h}{4 \sinh(q l)} \left[ \sinh(qz) + e^{-ql} \sinh(q(l-z)) - e^{-q z} q z - e^{q z} q(l-z) \right]  \nonumber \\
& +\frac{e^{-q(l-z)}}{2} \left[ -i f_{2x}^u(q,t) - f_{2z}^u(q,t) \right] + \frac{e^{-q z}}{2} \left[ -i f_{2x}^d(q,t) + f_{2z}^d(q,t) \right].
\end{align}

The flow field in the region $z>l$ is 
\begin{align}
v_x(q,z,t) & = \frac{e^{-q(z-l)}}{4 \eta _1 q}\left[ iq(l-z)f_{1z}(q,t)+(1+q(l-z))f_{1x}(q,t) \right],\\
v_z(q,z,t) & = \frac{e^{-q(z-l)}}{4 \eta _1 q}\left[ iq(l-z)f_{1x}(q,t)+(1-q(l-z))f_{1z}(q,t) \right],
\end{align}
and the flow field in the region $z<l$ is 
\begin{eqnarray}
v_x(q,z,t) &=& \frac{1}{4 \eta _2 q}\frac{i \chi q h(q,t)}{4 \sinh(q l)}
\left\{ (-1-qz  + q^2 z^2)\sinh(qz) + (q z - q^2 z^2) \cosh (q z)  \right.  \nonumber \\
 &&  - \left[e^{ql}(2q(l-z)- q^2(l-z)^2) + e^{-ql}(1-q(l-z)) \right]\sinh(q(l-z)) \nonumber \\
 && \left.
         - \left[-e^{ql}(2q(l-z)-q^2(l-z)^2) + e^{-ql} q(l-z) \right] \cosh(q(l-z)) 
\right\} \nonumber \\
&& + \frac{e^{-q (l-z)}}{4 \eta _2 q} \left[ iq(l-z) f_{2z}^{\rm u}(q,t) + (1-q(l-z) f_{2x}^{\rm u}u(q,t)  \right]
\nonumber \\
&& + \frac{e^{-q z}}{4 \eta _2 q} \left[-iq z f_{2z}^{\rm d}(q,t) + (1-q z) f_{2x}^{\rm d}d(q,t)  \right],
\end{eqnarray}
\begin{eqnarray}
v_z(q,z,t) &=& -\frac{1}{4 \eta_2 q}\left( \frac{\chi q h(q,t)}{4 \sinh(q l)} \right)
\left\{ (1+q z + q^2 z^2) \sinh (q z) - (q z + q^2 z^2) \cosh (q z)  \right.  \nonumber \\
 &&  - \left[e^{ql}q^2 (l-z)^2) + e^{-ql}(1+q(l-z)) \right]\sinh(q(l-z)) \nonumber \\
 && \left.
         + \left[e^{ql} q^2 (l-z)^2 + e^{-ql} q(l-z) \right] \cosh(q(l-z)) 
\right\} \nonumber \\
&& + \frac{e^{-q (l-z)}}{4 \eta _2 q} \left[ iq(l-z) f_{2x}^{\rm u}(q,t) + (1+q(l-z) f_{2z}^{\rm u}(q,t)  \right]
\nonumber \\
&& + \frac{e^{-q z}}{4 \eta _2 q} \left[-iq z f_{2x}^{\rm d}(q,t) + (1+q z) f_{2z}^{\rm d}(q,t)  \right].
\end{eqnarray}

The boundary forces are solved by imposing the boundary conditions.  
From the continuity of flow field at the membrane, we have 
\begin{eqnarray}
\eta _2 f_{1x}(q,t) &=& \eta _1 \left\{
\frac{i \chi q h(q, t)}{4 \sinh (q l)} \left[
     (-1-ql + q^2 l^2) \sinh(q l ) + ( q l - q^2 l^2) \cosh(q l) 
                                                                   \right] 
\right. \nonumber \\
&& \left. +f_{2x}^{\rm u}(q,t) + e^{-q l } \left[ (1-q l) f_{2x}^{\rm d}d (q, t) - i q l f_{2z}^{\rm d}(q , t) \right]
\right\},
\label{eq:bc1}
\end{eqnarray}
\begin{eqnarray}
\eta _2 f_{1z}(q,t) &=&  - \eta _1 \left\{
\frac{ \chi q h(q, t)}{4 \sinh (q l)} \left[
     (1+ ql + q^2 l^2) \sinh(q l ) - ( q l + q^2 l^2) \cosh(q l) 
                                                                   \right] 
\right. \nonumber \\
&& \left. +f_{2z}^{\rm u}(q,t) + e^{-q l } \left[ (1+q l) f_{2z}^{\rm d} (q, t) - i q l f_{2x}^{\rm d}(q , t) \right]
\right\}.
\label{eq:bc2}
\end{eqnarray}
From the no slip condition at $z=0$, we have 
\begin{eqnarray}
f_{2x}^d (q,t)&=& \frac{i \chi q h(q,t)}{4 \sinh(q l)}
 \left\{ ((1 - q l)e^{-ql}+(2 q l - q^2 l^2)e^{ql})\sinh (q l) \right. \nonumber \\
         & &\left. + (q l e^{- q l } + (-2 q l + q^2 l^2)e^{ql})\cosh(q l) \right\} \nonumber \\
& & - e^{-q l}\left[ iq l f_{2z}^{\rm u}(q,t) + (1-q l) f_{2x}^{\rm u}(q, t))\right],
\label{eq:bc3}
\end{eqnarray}
\begin{eqnarray}
f_{2z}^d(q,t) &=& \frac{\chi q h(q,t)}{4 \sinh(q l)} \left[ 
  (q^2 l^2 e^{q l } + q l e^{-q l})\cosh (q l ) 
  - (q^2 l ^2 e^{q l} + (1+ q l )e^{-q l })\sinh(q l )
\right] \nonumber \\
& & - e^{- q l } \left[ 
(1+q l ) f_{2z}^{\rm u}(q,t) + i q l f_{2x}^{\rm u}(q, t)
\right].
\label{eq:bc4}
\end{eqnarray}

The continuity of shear stress across the membrane leads to 
\begin{eqnarray}
 f_{1x}(q,t) &=& -\frac{1}{4} \left\{
 \frac{i \chi q h(q,t)}{4 \sinh(q,t)} \left[(1+q l - q^2 l^2)\sinh(q l) - (q l - q ^2 l^2)\cosh(ql)
 \right] \right. \nonumber \\
 && \left. + f_{2x}^{\rm u}(q,t) + e^{-q l} \left[ 
 i q l f_{2z}^{\rm d}(q,t) + (-1 + q l ) f_{2x}^{\rm d}(q,t)
 \right]
 \right\},
 \label{eq:bc5}
\end{eqnarray}
and the generalized Laplace condition for the normal stress across the membrane gives
\begin{eqnarray}
- \frac{\delta F}{\delta h(-q,t)} &=& \frac{1}{2}\frac{\chi q h(q,t)}{4 \sinh (q l)} \left[
(1+ql + q^2l^2)\sinh(q l) - (ql+q^2 l^2)\cosh(q l ) \right] \nonumber \\
&& 
+ \frac{f_{1z}(q,t)}{2} + \frac{f_{2z}^{\rm u}(q, t)}{2} + e^{- q l } 
\left( \frac{i q l }{2}f_{2x}^{\rm d}(q, t)- \frac{1+q l }{2}f_{2z}^{\rm d}(q, t) \right).
\label{eq:bc6}
\end{eqnarray}
After solving Eqs.~(\ref{eq:bc1}), (\ref{eq:bc2}), (\ref{eq:bc3}), (\ref{eq:bc4}), (\ref{eq:bc5}) and (\ref{eq:bc6}), 
the kinematic condition $\partial h/\partial t = v_z|_{z=l}$ gives the evolution equation~(\ref{eq:dhdt}) for membrane height.

\section{Approximate solution of membrane dynamics at $q l \ll 1$}

In this Appendix, we present the calculation for  membrane dynamics in the limit $q l \ll 1$.  
To be consistent and to compare to the $q l \ll 1$ limit provided by Eq.~(\ref{eq:smallq}), we mark the magnitude 
of the sub-leading terms in each step of our calculation, and only keep the leading terms.  This also helps us to compare our result to the calculation based on lubrication approximation~\cite{Ramaswamy,Basu}.

Let the typical magnitude of $v_x$ in the active fluid be $U$.
Then the incompressibility condition gives the typical magnitude of $v_z$ in the active fluid as 
\begin{eqnarray}
v_z \sim q l U,~~~~~z < l.
\end{eqnarray}
In this limit, the angle of the director field is 
\begin{eqnarray}
\theta (q, z, t) = i q h(q, t)\frac{\sinh (q z)}{\sinh (q l)} \approx i q  h(q, t) \frac{z}{l}.
\end{eqnarray}
By using $\partial _z^2 \sim 1/l^2 \gg q ^2 \sim \partial _x^2$, the momentum equation for the active fluid can be approximated as
\begin{eqnarray}
\eta _2 \partial _z^2 v_x - \partial _x p + \frac{i \chi q h}{2l}+\mathcal{O}(q^2l^2) = 0,
\label{eq:thin_film_x_momentum}
\end{eqnarray}
and 
\begin{eqnarray}
-\partial _z p - \frac{\chi q ^2  h}{2l}z+ \mathcal{O}(q^2l^2)=0.
\end{eqnarray}
Here $\mathcal{O}(q^2l^2)$ means that terms that are dropped are of order $q ^2 l ^2$ or smaller compared to the leading terms.

At $z=0$, the no slip condition applies.  The continuity of shear stress at $z=l$ leads to 
\begin{eqnarray}
\eta _2 \partial _z v_x - \frac{i \chi q h}{2 l }z + \mathcal{O}(E q l)= 0,
\label{eq:thin_film_shear}
\end{eqnarray}
where $\mathcal{O}(E q l)$ means that the term that is dropped comes from shear stress of the passive fluid, 
it is of order $q l$ compare to the leading terms.  
This is because the passive fluid above the membrane satisfies $\partial _z v_x \sim q v_x$, and the continuity of flow field at $z=x$ means that $\eta _1 (\partial _z v_x + \partial _x v_z)|_{z=l^+} \sim E q l \ \eta _2 (\partial _z v_x+\partial _x v_z)|_{z=l_-} $.  
Similar analysis for the normal stress across the membrane leads to 
\begin{eqnarray}
- \frac{\partial F}{\partial h}= -p+ \mathcal{O}(q^2 l^2).
\end{eqnarray}  
From the momentum equation, the typical magnitude of pressure in the active fluid satisfies $q p \sim \eta _2 U/l^2$, 
leading to $p \sim \eta _2 U / (q l^2)$ in the active fluid.  
From the momentum equation in the passive fluid, the pressure in the passive fluid is of order $q ^2 l^2$  
compared to the pressure in the active fluid. 
The viscous stress in the active fluid is $\eta _2 \partial _z v_z \sim \eta _2 U q $.
From incompressibility and continuity of flow field across the membrane, the typical magnitude of flow in the passive fluid is $v_z \sim v_x \sim U$. 
This gives us $\eta _1\partial _z v_z \sim \eta _1 q U$.  
Therefore the viscous stress in both active and passive fluids, like the pressure in the passive fluid, is of order $q^2 l^2$ compared to the pressure in the active fluid.

To obtain the evolution equation for the membrane height, we first integrate the $z$-component of the momentum equation from $z$ to $l$ and use the boundary condition for normal stress. 
After that the pressure in the active fluid is solved as 
\begin{eqnarray}
p(q,z,t) =  \frac{\partial F}{\partial h(-q,t)}+\frac{\chi}{4} \frac{q^2(l^2-z^2)}{l}h(q,t)+\mathcal{O}(q^2 l^2).
\label{eq:thin_film_p}
\end{eqnarray}
Next, we substitute pressure into the $x$-component of the momentum equation, integrate over $z$ two times, use the continuity of shear stress at $z=l$ and no slip condition at $z=0$.
Then $v_x$ in the active fluid becomes 
\begin{eqnarray}
v_x &=& -\frac{1}{\eta _2} \left[
\left( i q \frac{\partial F}{\partial h(q,t)}-i\frac{\chi qh(q,t)}{2l} \right)\left(l z- \frac{z^2}{2}\right)
+ \frac{i \chi q h(q,t)}{2}z +\mathcal{O}(E q l) \right].
\label{eq:thin_film}
\end{eqnarray}
Note that, first, the dominant contribution from the active stress to $v_x$ comes from the active stress in the shear stress continuity condition Eq.~(\ref{eq:thin_film_shear}) and $i\chi q h /(2l)$ term in the momentum equation~(\ref{eq:thin_film_x_momentum}).  The contribution from active stress term in the pressure in Eq.~(\ref{eq:thin_film_p}) is of order $q^2 l^2$ compared to these two terms.  Second, the ``leading dropped term'' is $\mathcal{O}(E q l)$, which comes from the viscous stress of the passive fluid that has been neglected in the shear stress continuity boundary condition Eq.~(\ref{eq:thin_film_shear}).

Finally, using the kinematic boundary condition $\partial _t h = v_z|_{z=l}$ and incompressibility condition, we obtain the evolution equation of the membrane height in the $ql \ll 1$ limit as 
\begin{eqnarray}
\frac{\partial h(q,t)}{\partial t} &=& v_z(q,l,t) \nonumber \\
&=& \int _0^l \partial _z v_z (q,z,t) dz \nonumber \\
&=& -iq \int _0^l v_x (q,z,t)dz \nonumber \\
&=& - \frac{1}{\eta_2}\left[ 
(\gamma q + \kappa q^3)\frac{q^3l^3}{3}+ \chi \frac{q^2l^2}{12}  + \mathcal{O}(E q l)
\right] h(q,t).
\end{eqnarray}
This gives the leading terms in Eq.~(\ref{eq:smallq}).  The leading correction is of order $q l$ compared to the leading 
term, and it comes from the shear stress of the passive fluid.

It is interesting to compare our approach with the lubrication approximation presented in Refs.~\cite{Ramaswamy,Basu}.  
Although Refs.~\cite{Ramaswamy,Basu} considered active fluid with polar order and fluctuating active particle density, 
when the in-plane ($xy$) fluctuation of the director field and density fluctuations are neglected, their model is basically identical to our model. 
However, in Refs.~\cite{Ramaswamy,Basu}, they neglected the contribution of active stress to the shear stress at $z=l$.  
This makes the leading active term in the resulting membrane equation different from Eq.~(\ref{eq:thin_film}).  Furthermore, according to our calculation, terms of order $q l$ or smaller compared to the leading terms can come from the passive fluid at $z>l$, or $v_z$ in the active fluid.  These are neglected in lubrication approximation.  Therefore the sub-leading terms in the membrane dynamic equation of  Refs.~\cite{Ramaswamy,Basu} are also different from Eq.~(\ref{eq:smallq}).  Because of these differences, 
they predicted a long-wavelength instability of the system against director splay in the $xz$ plane, but our systematic thin film calculation predicts that the system is stable in the long wavelength limit.

\end{document}